\documentclass[twocolumn,tighten]{aastex62}
\usepackage{natbib}
\usepackage{epsfig}
\usepackage{xcolor}
\usepackage[all]{hypcap} 

\newcommand{\acounits}{\mbox{M$_\odot$ pc$^{-2}$ (K km s$^{-1}$)$^{-1}$}}

\newcommand{\ARI}{\affiliation{Astronomisches Rechen-Institut, Zentrum f\"ur Astronomie der Universit\"at Heidelberg, M\"onchhofstra{\ss}e 12-14, D-69120 Heidelberg, Germany}}

\newcommand{\ESO}{\affiliation{European Southern Observatory, Karl-Schwarzschild-Stra{\ss}e 2, D-85748 Garching bei M\"unchen, Germany}}

\newcommand{\ITA}{\affiliation{Institute f\"ur theoretische Astrophysik, Zentrum f\"ur Astronomie der Universit\"at Heidelberg, Albert-Ueberle Str. 2, 69120 Heidelberg, Germany}}

\newcommand{\MPIA}{\affiliation{Max Planck Institute for Astronomy, K\"onigstuhl 17, 69117, Heidelberg, Germany}}

\newcommand{\NRAO}{\affiliation{National Radio Astronomy Observatory, 520 Edgemont Road, Charlottesville, VA 22903-2475, USA}}

\newcommand{\OAN}{\affiliation{Observatorio Astron\'omico Nacional (IGN), C/ Alfonso XII, 3, 28014 Madrid, Spain}}

\newcommand{\OSU}{\affiliation{Department of Astronomy, The Ohio State University\\4055 McPherson Laboratory, 140 West 18th Ave, Columbus, OH 43210, USA}}

\newcommand{\MPE}{\affiliation{Max-Planck-Institut f\"ur extraterrestrische Physik, Giessenbachstra{\ss}e 1, 85748 Garching, Germany}}

\shorttitle{Comparing Dense Gas Fractions to Cloud-Scale Mean Densities}
\shortauthors{Gallagher et al.}

\begin{document}

\title{Do Spectroscopic Dense Gas Fractions Track Molecular Cloud Surface Densities?}

\correspondingauthor{Molly J. Gallagher}
\email{gallagher.674@osu.edu}

\author{Molly J. Gallagher}
\OSU

\author{Adam K. Leroy}
\OSU

\author{Frank Bigiel}
\affiliation{Argelander-Institut f\"ur Astronomie, Universit\"at Bonn, Auf dem H\"ugel 71, 53121 Bonn, Germany}

\author{Diane Cormier}
\affiliation{AIM, CEA, CNRS, Universit\'e Paris-Saclay, Universit\'e Paris Diderot, Sorbonne Paris Cit\'e, F-91191 Gif-sur-Yvette, France}

\author{Mar\'{i}a J. Jim\'{e}nez-Donaire}
\affiliation{Harvard-Smithsonian Center for Astrophysics, 60 Garden St., Cambridge, MA 02138, USA}

\author{Annie Hughes}
\affiliation{CNRS, IRAP, 9 av. du Colonel Roche, BP 44346, F-31028 Toulouse cedex 4, France}
\affiliation{Universit\'{e} de Toulouse, UPS-OMP, IRAP, F-31028 Toulouse cedex 4, France}

\author{J\'{e}r\^{o}me Pety}
\affiliation{Institut de Radioastronomie Millim\`{e}trique (IRAM), 300 Rue de la Piscine, F-38406 Saint Martin d'H\`{e}res, France}
\affiliation{Observatoire de Paris, 61 Avenue de l'Observatoire, F-75014 Paris, France}

\author{Eva Schinnerer}
\MPIA

\author{Jiayi Sun}
\OSU

\author{Antonio Usero}
\OAN

\author{Dyas Utomo}
\OSU

\author{Alberto Bolatto}
\affiliation{Department of Astronomy, University of Maryland, College Park, MD 20742-2421, USA}

\author{M\'{e}lanie Chevance}
\ARI

\author{Christopher M. Faesi}
\MPIA

\author{Simon C. O. Glover}
\ITA

\author{Amanda A. Kepley}
\NRAO

\author{J. M. Diederik Kruijssen}
\ARI

\author{Mark R. Krumholz}
\affiliation{Research School of Astronomy \& Astrophysics, Australian National University, Canberra, ACT 2611, Australia}
\affiliation{Centre of Excellence for Astronomy in Three Dimensions (ASTRO-3D), Australia}

\author{Sharon E. Meidt}
\MPIA

\author{David S. Meier}
\affiliation{Department of Physics, New Mexico Institute of Mining and Technology, 801 Leroy Place, Soccoro, NM 87801, USA}
\affiliation{National Radio Astronomy Observatory, P. O. Box O, 1003 Lopezville Road, Socorro, NM, 87801, USA}

\author{Eric J. Murphy}
\NRAO

\author{Miguel Querejeta}
\ESO
\OAN

\author{Erik Rosolowsky}
\affiliation{Department of Physics, University of Alberta, Edmonton, AB T6G 2E1, Canada}

\author{Toshiki Saito}
\MPIA

\author{Andreas Schruba}
\MPE

\begin{abstract}
We use ALMA and IRAM 30-m telescope data to investigate the relationship between the spectroscopically-traced dense gas fraction and the cloud-scale (120~pc) molecular gas surface density in five nearby, star-forming galaxies. We estimate the dense gas mass fraction at 650~pc and 2800~pc scales using the ratio of \mbox{HCN~(1-0)} to \mbox{CO~(1-0)} emission. We then use high resolution (120~pc) \mbox{CO~(2-1)} maps to calculate the mass-weighted average molecular gas surface density within 650~pc or 2770~pc beam where the dense gas fraction is estimated. On average, the dense gas fraction correlates with the mass-weighted average molecular gas surface density. Thus, parts of a galaxy with higher mean cloud-scale gas surface density also appear to have a larger fraction of dense gas. The normalization and slope of the correlation do vary from galaxy to galaxy and with the size of the regions studied. This correlation is consistent with a scenario where the large-scale environment sets the gas volume density distribution, and this distribution manifests in both the cloud-scale surface density and the dense gas mass fraction. 
\end{abstract}


\section{Introduction}
\label{sec:Introduction}
    
Star formation is tied to the presence of dense gas. We observe this in the Milky Way, where stars form primarily in dense substructures within molecular clouds \citep[e.g.,][]{LADA03, LADA10, LADA12, HEIDERMAN10, EVANS14, ANDRE14}. We also observe this in external galaxies. There the dense gas mass, traced by high effective critical density molecular lines, correlates with the star formation rate \citep[e.g., see][]{GAO04, GARCIABURILLO12, USERO15, CHEN15, BIGIEL16, GALLAGHER18, KEPLEY18}. Both Galactic and extragalactic observations also indicate that gas volume density and its relationship to star formation change as a function of environment \citep[e.g.,][]{GAO04, LONGMORE13, KRUIJSSEN14, USERO15,BIGIEL16,GALLAGHER18}. Measuring the gas volume density across many environments is key to understand what drives these variations.

Observers use two main methods to gauge the distribution of volume densities in the  molecular interstellar medium (ISM) of other galaxies. First, one can estimate the surface density of molecular clouds by imaging of a line that traces the bulk molecular gas mass, such as CO emission \citep[e.g.,][]{HUGHES13,COLOMBO14}. Given an estimate of the line-of-sight depth, e.g. from a cloud size or adopted scale height, we can convert this surface density to a volume density. This method requires high physical resolution ($\lesssim 100$~pc) to avoid bias from beam dilution.

One can also infer the distribution of gas volume density\footnote{Unless otherwise stated, "density" refers to gas volume density throughout the paper.} from observations of multiple molecular emission lines that are excited at different effective critical densities, $n_{\rm eff}$ \citep[e.g.,][]{GAO04,GARCIABURILLO12}. To first order, the luminosity of a line traces the mass of gas above its $n_{\rm eff}$. Changes in the ratio of intensities between lines with different $n_{\rm eff}$ can indicate a changing ratio in the mass above each density \citep[though there are subtleties, see][]{KRUMHOLZ07, LEROY17}. This method constrains the volume density distribution within the beam without the need to resolve individual clouds. In the simple case of a bulk gas tracer (here we use CO~(1-0) with $n_{\rm eff} \approx 1\times 10^2$~cm$^{-3}$) and a dense gas tracer \citep[here we use HCN~(1-0) with $n_{\rm eff} \approx 5\times10^3$~cm$^{-3}$, e.g.,][]{ONUS18}, this method traces the dense gas fraction ($f_{dense}$). Because high effective critical density lines also tend to be faint, this method has been mostly employed in low resolution, high sensitivity data sets, which average over scales much larger than that of an individual cloud \citep[e.g.,][]{GAO04, USERO15, CHEN15, BIGIEL16, GALLAGHER18}.

These two methods trace density in different ways at different scales but they should be related. If the mean volume density (and thus also the mean surface density) of a cloud increases or decreases, then we might expect the fraction of gas above some effective density (e.g., that of HCN) to rise or drop in parallel. Exactly this prediction arises from turbulent cloud models \citep[e.g.,][]{PADOAN02, KRUMHOLZ07, FEDERRATH13}. In these models, increasing the mean volume density of a cloud ``slides'' the density distribution to a higher range of values and thus should increase $f_{dense}$. This work represents the first observational test of this correlation.

In this Letter, we leverage the results of recent observing campaigns using the Institute for Radio Astronomy in the Millimeter range (IRAM) \mbox{30-m}\footnote{This work is partially based on observations carried out with the IRAM \mbox{30-m} telescope. IRAM is supported by INSU/CNRS (France), MPG (Germany) and IGN (Spain).} and the Atacama Large Millimeter/submillimeter Array (ALMA) to compare these two density estimates. 
We estimate the cloud-scale molecular gas surface density from the PHANGS-ALMA survey\footnote{\url{http://phangs.org}} (P.I.\ E.~Schinnerer; A.~K.~Leroy et~al. in preparation). PHANGS-ALMA is mapping CO~(2-1) at high resolution (${\sim}100$~pc) across 74 nearby galaxies. We compare this to $f_{dense}$ estimated using HCN~(1-0) and CO~(1-0) (hereafter referred to as HCN and CO, respectively) maps from ALMA \citep{GALLAGHER18} and the IRAM Large Program EMPIRE \citep[][Jimenez Donaire et~al.\ in prep.]{BIGIEL16}. 

Section \ref{sec:data} summarizes how we calculate the HCN/CO ratio (\S \ref{subsec:hcn-to-co}), the characteristic cloud-scale CO intensity with a larger $\sim$kpc-scale aperture (\S \ref{subsec:avgcssd}), and the mean relation between the two (\S \ref{subsec:BD}). Section \ref{sec:results} presents the observed correlation (\S \ref{subsec:OC}) and discusses its physical implications (\S \ref{subsec:TI}). We compare our results to simple density distribution models (\S \ref{subsec:SM}). Section \ref{sec:summary} summarizes our findings.

\section{Data and Methods}
\label{sec:data}

\capstartfalse
\begin{deluxetable*}{cccccc}[ht!]
\tabletypesize{\scriptsize}
\tablecaption{Galaxy Sample\label{tab:galaxy_data}}
\tablewidth{0pt}
\tablehead{
\colhead{Galaxy} & 
\colhead{HCN Survey} &
\colhead{High Res.~CO} &
\colhead{Distance} &
\colhead{$8''$ Resolution} &
\colhead{$34''$ Resolution} \\[2px]
 \multicolumn{3}{c}{}  & \multicolumn{1}{c}{(Mpc)} & \multicolumn{1}{c}{(kpc)} & \multicolumn{1}{c}{(kpc)}  
} 
\startdata
NGC 3351 & ALMA & PHANGS-ALMA & 10.0 & 0.39 & \nodata \\ 
NGC 3627 & ALMA, EMPIRE & PHANGS-ALMA & 8.28 & 0.32 & 1.36\\ 
NGC 4254 & ALMA, EMPIRE & PHANGS-ALMA & 16.8 & 0.65 & 2.77  \\ 
NGC 4321 & ALMA, EMPIRE & PHANGS-ALMA & 15.2 & 0.59 & 2.51  \\ 
NGC 5194 &  EMPIRE & PAWS & 8.39 & \nodata & 1.38
\enddata
\tablecomments{{\bf HCN Data}: ALMA---\citet{GALLAGHER18}, EMPIRE---IRAM \mbox{30-m} Large Program \citep[][Jimenez-Donaire et~al.\ in prep.]{BIGIEL16}. {\bf High Res.\ CO Data}: PAWS CO~(1-0)---\citet{SCHINNERER13}. PHANGS-ALMA CO~(2-1)---Leroy et~al.\ in prep. and \citet{SUN18}. {\bf Distance}: adopted distance in Mpc from the Extragalactic Distance Database \citep{TULLY09}. {\bf 8$''$ (34$''$) Resolution}: physical resolution corresponding to $8''$ and $34''$ at our adopted distances.}
\end{deluxetable*}
\capstarttrue

Table~\ref{tab:galaxy_data} lists our targets. We consider two samples, based on the availability of \mbox{HCN~(1-0)} data, which we analyze separately. The EMPIRE sample has $34''$ resolution HCN maps from the IRAM \mbox{30-m}. The ALMA sample has $8''$ HCN maps. NGC~3627, NGC~4254, and NGC~4321 appear in both samples. NGC~3351 appears only in the ALMA sample and NGC~5194 appears only in the EMPIRE sample.

\subsection{HCN~(1-0) to CO~(1-0) Ratio}
\label{subsec:hcn-to-co}

\textbf{EMPIRE HCN Data:} EMPIRE \citep{BIGIEL16} used the IRAM \mbox{30-m} telescope to map HCN emission from nine galaxies, four of which have high resolution CO maps suitable for our experiment. The EMPIRE maps cover the whole star-forming disk of each galaxy, but have relatively poor ($34''$) angular resolution. 

\begin{figure*}[th!]
\begin{center}
\includegraphics[height=2.25in]{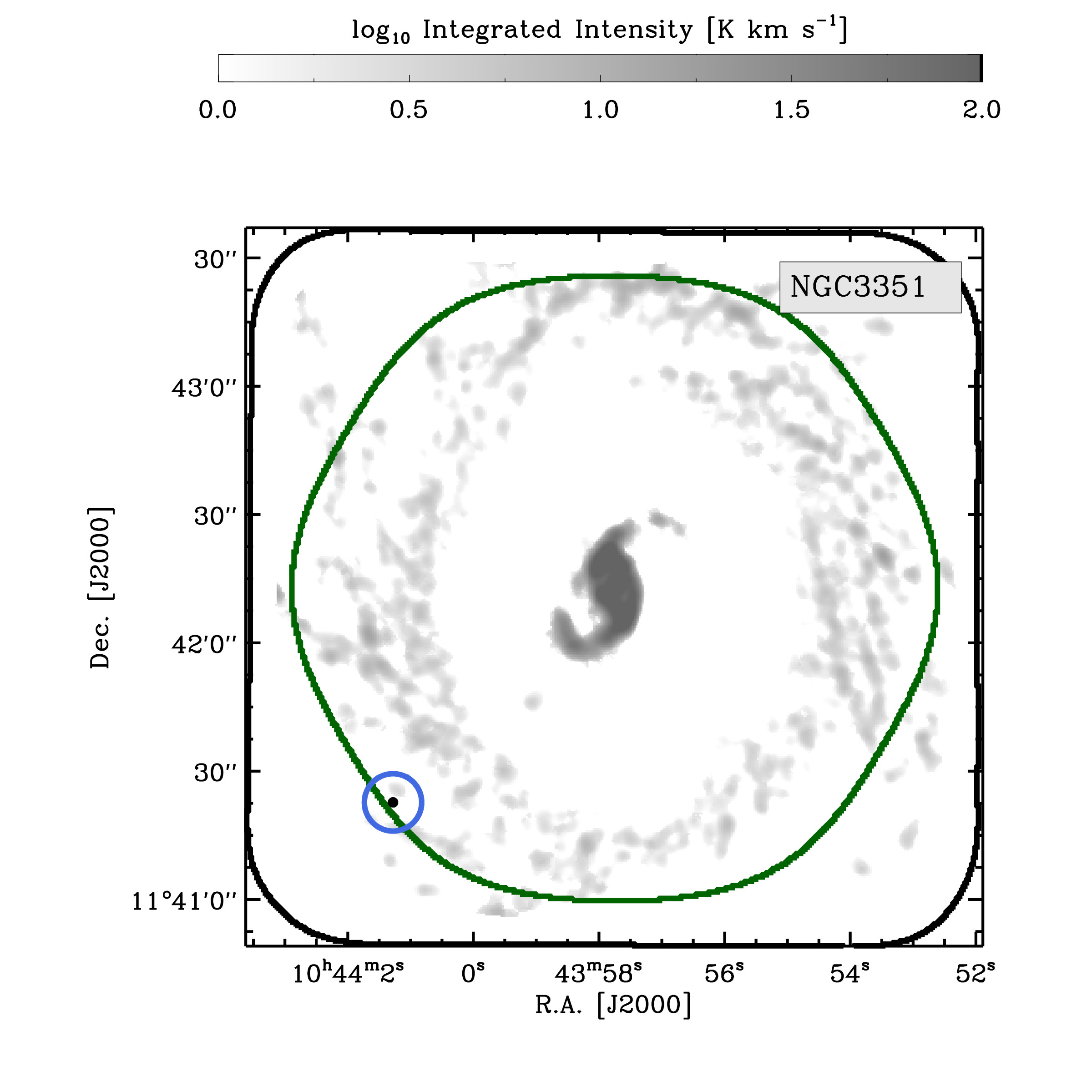}
\includegraphics[height=2.25in]{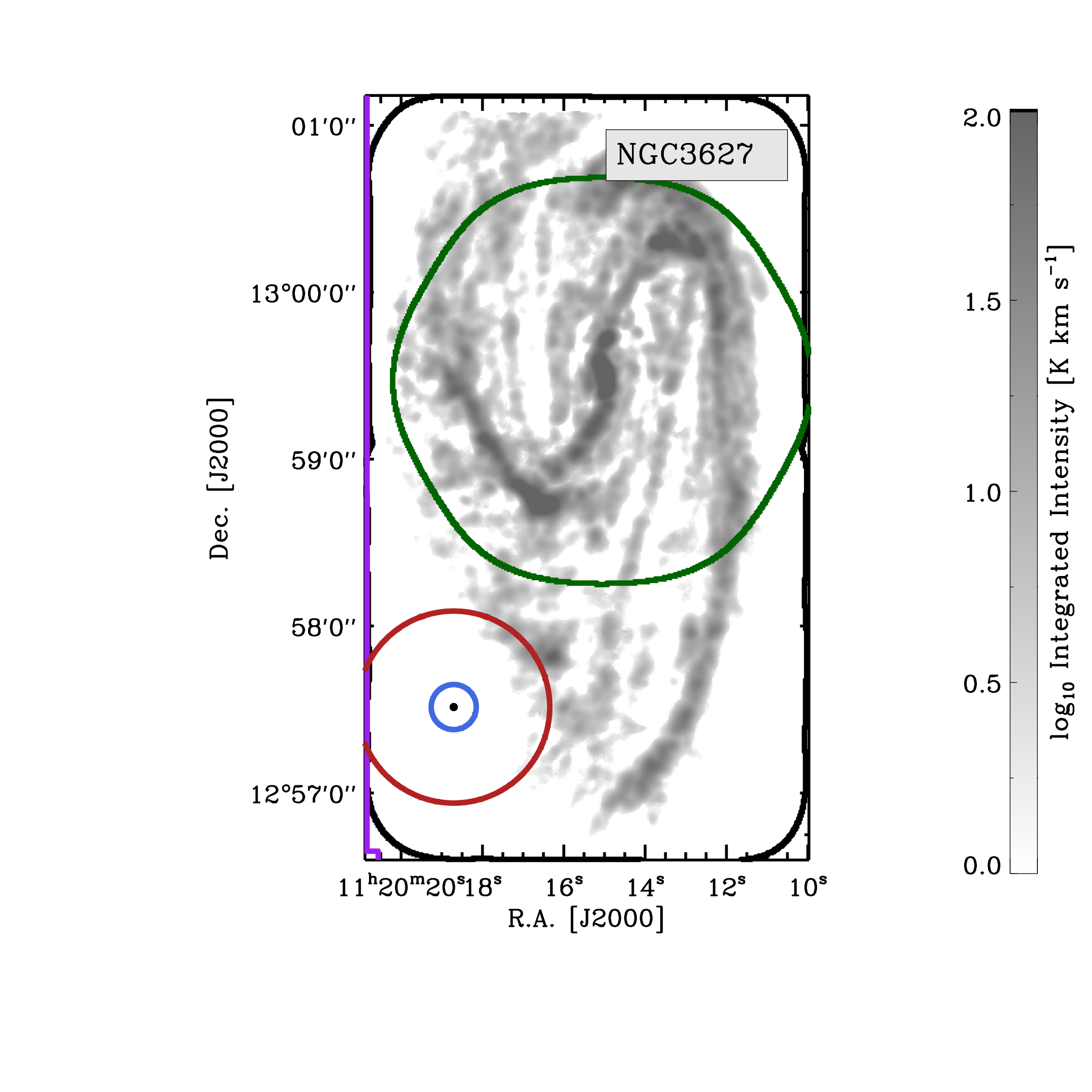}
\includegraphics[height=2.25in]{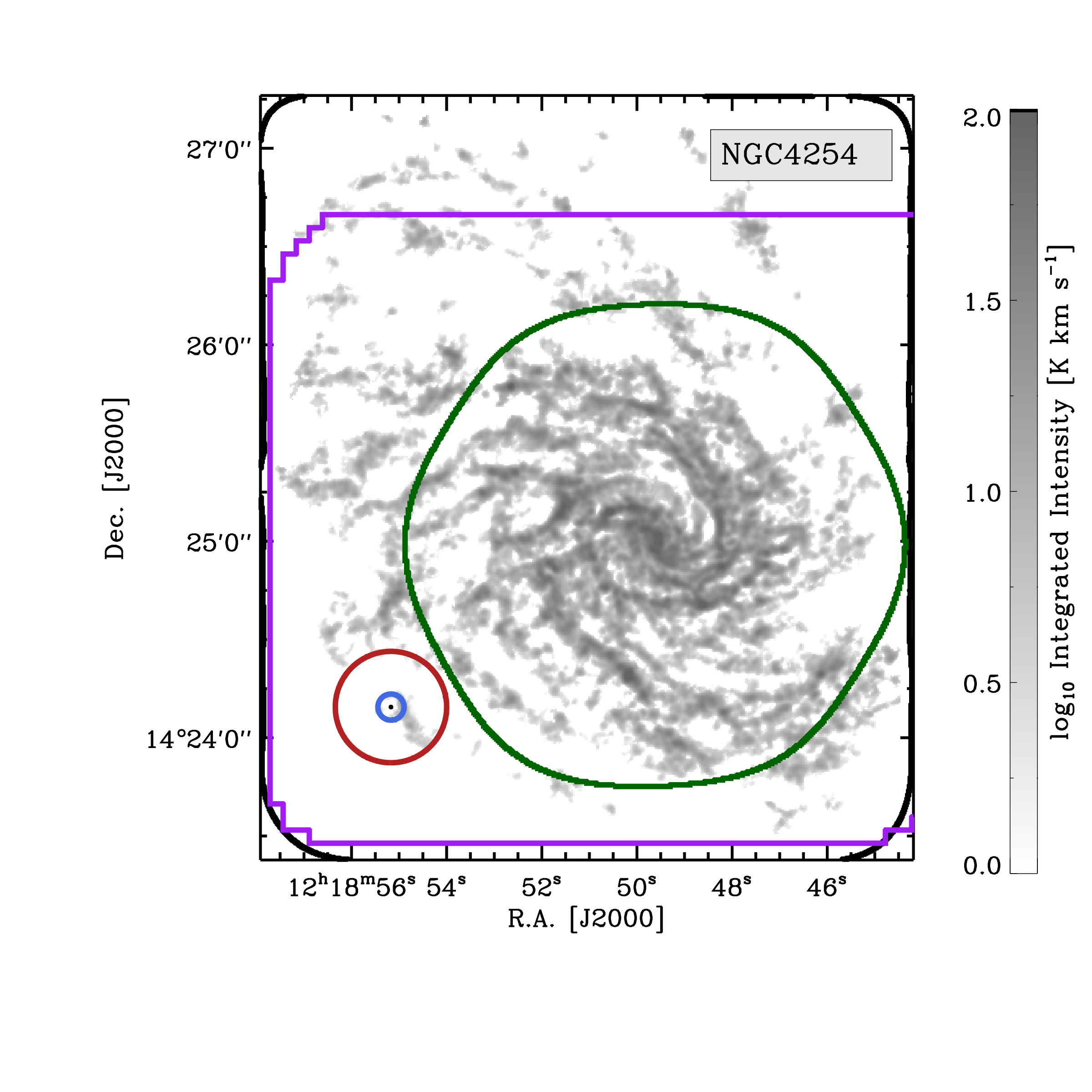}
\includegraphics[height=2.25in]{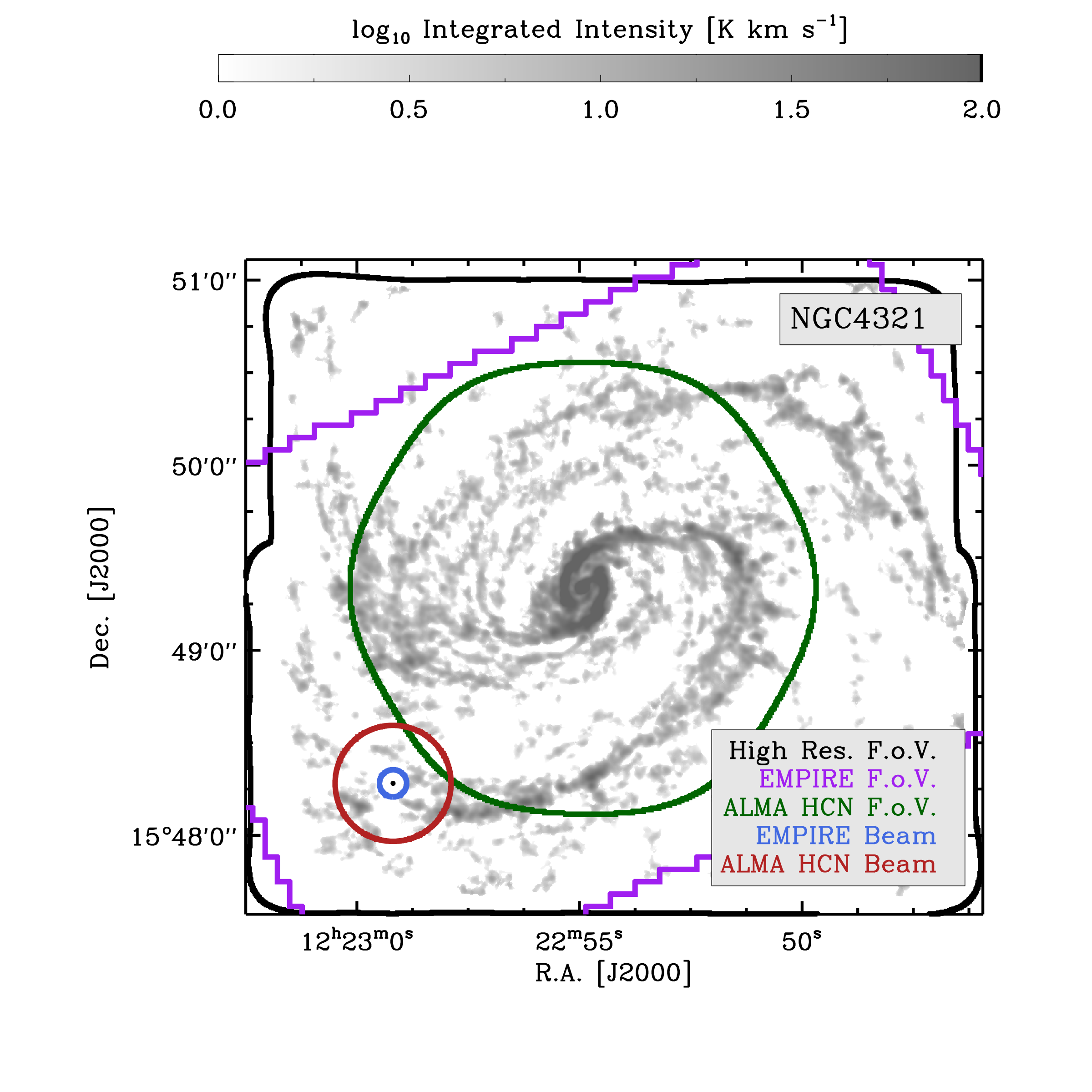}
\includegraphics[height=2.25in]{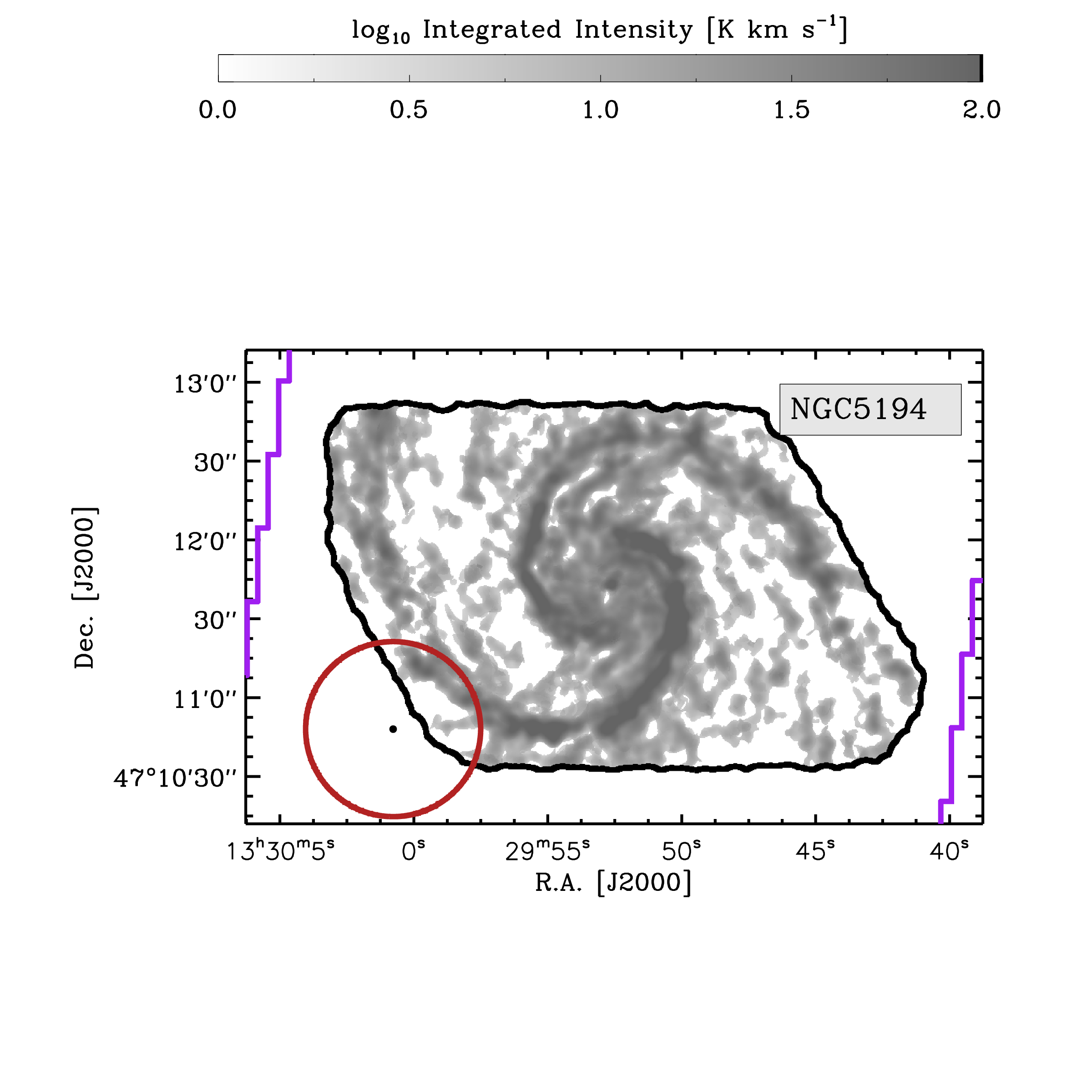}
\caption{Integrated intensity maps at 120~pc resolution of CO~(2-1) emission for our PHANGS-ALMA targets and CO~(1-0) for NGC~5194 (grayscale). In the lower left corner of each map, circles represent the three resolutions used in this work: the 2770~pc common resolution of the EMPIRE HCN data (red), the 650~pc common resolution of the ALMA HCN data (blue), and the 120~pc common resolution of the cloud scale CO maps (black). The thick black lines show the field of view of the high resolution map, and the green and purple lines represent the fields of view of the ALMA (green) and EMPIRE (purple) HCN data.
\label{fig:fov}}
\end{center}
\end{figure*}

Jimenez Donaire et~al. (in prep.) present a full description of EMPIRE, including the new \mbox{CO~(1-0)} maps \citep[see also][]{CORMIER18}, which we pair with the HCN maps to measure the \mbox{HCN~(1-0)} to \mbox{CO~(1-0)} ratio. Briefly, EMPIRE uses the IRAM 30-m in on-the-fly mapping mode to cover the area of active star formation in each target (see Figure \ref{fig:fov}). Observations were conducted from 2012 to 2016. The data were calibrated using GILDAS, extracted at $4$~km~s$^{-1}$ spectral resolution and then further reduced using an in-house pipeline. The pipeline fits and subtracts a second order polynomial baseline, avoiding regions of the spectrum known to have bright CO emission. It rejects spectra with measured noise significantly larger than that predicted by the radiometer equation. Then it projects the data on to grids with pixel size of $4~\arcsec$. The adopted gridding kernel convolved with the IRAM~30-m beam yields a final angular resolution of $34\arcsec$. Finally, the pipeline converts the data to main beam temperature units, assuming main beam and forward efficiencies of 0.78 and 0.94 (Kramer, Penalver, \& Greve 2013, IRAM calibration papers).

\textbf{ALMA HCN Data:} \citet{GALLAGHER18} mapped HCN emission from four galaxies and assembled matched-resolution \mbox{CO~(1-0)} maps from the literature\footnote{The matched-resolution CO~(1-0) maps come from ALMA, CARMA, and BIMA. All include short and zero spacing data. \citet{GALLAGHER18} give more details.} The ALMA maps cover out to $r_{\rm gal} = 3.5{-}6$~kpc, a smaller area than EMPIRE (Figure \ref{fig:fov}) but have a sharper $8''$ resolution.

We take CO observations from BIMA SONG (\citealt{HELFER03}) for NGC3351 and NGC3627. These cubes include data from both the BIMA interferometer and short-spacing data from the NRAO 12m single dish telescope on Kitt Peak. We take interferometric CO observations from CARMA STING (\citealt{RAHMAN11}) for NGC4254. We combine this with single dish data from the CO extension to the IRAM EMPIRE survey \citep[][M. Jim\'enez-Donaire et al. in prepation]{CORMIER18}. We take CO data from the ALMA science verification program for NGC4321. This includes both main 12-m array and Atacama Compact Array (ACA) short spacing and total power data. We take the CO data from EMPIRE for NGC5194.

ALMA observed HCN~(1-0) using a seven-field mosaic centered on the nucleus of the galaxy. The data were reduced using the CASA \citep{MCMULLIN07} package and  observatory-provided calibration scripts. They were then imaged using natural weighting with a small $u-v$ taper and a velocity resolution of 10~km~s$^{-1}$. Using the CASA task {\tt feather}, the ALMA cubes were combined with the IRAM-30m maps (mostly from EMPIRE) to correct for missing short spacing data. The resulting cubes were convolved to a resolution of $8\arcsec$, chosen to match archival CO and infrared data \citep[see][]{GALLAGHER18}. The statistical noise in each $10$~km~s$^{-1}$ channel is $\sim 5{-}10$~mK.

For each sample, we convolve the HCN~(1-0) and CO~(1-0) data from the literature for all targets to a common physical resolution set by the physical beam size at the most distant target. This is 650~pc for ALMA and 2770~pc for EMPIRE. At this common resolution, we measure HCN/CO, the ratio of the HCN(1-0) to CO(1-0) integrated intensities. We constructed integrated intensity map of HCN using a mask defined in position-position-velocity space from the CO~(1-0) cube. We then measure HCN/CO everywhere withing the fields of view that CO is detected \citep[see][]{GALLAGHER18}.

\begin{figure*}[th!]
\plotone{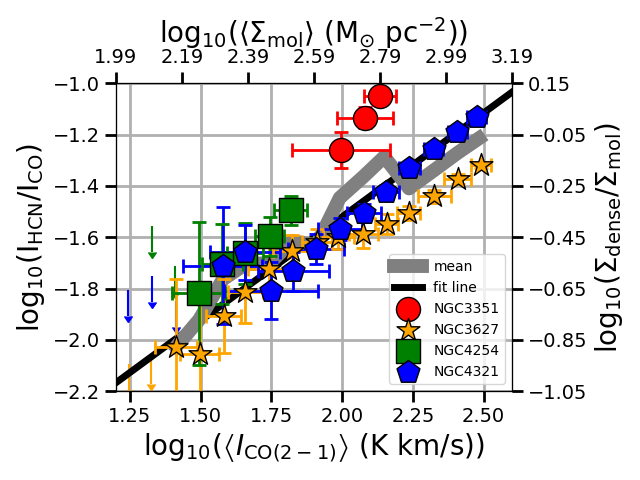}
\caption{\mbox{HCN~(1-0)/CO~(1-0)}, a spectroscopic tracer of $f_{dense}$, as a function of the molecular cloud surface density averaged within $\sim0.6$~kpc regions across our ALMA sample. The mean molecular cloud surface density is estimated from the mean cloud-scale CO~(2-1) intensity, $\left< I_{\rm CO(2-1)} \right>$, inside each $\sim 0.6$~kpc region. Colored points show mean HCN/CO in bins of fixed $\left< I_{\rm CO(2-1)} \right>$ for individual galaxies. The error bars on these points represent the average relative error for each bin. The gray line shows the mean HCN/CO at a given $\left< I_{\rm CO(2-1)} \right>$ combining all galaxies and weighting each galaxy equally. The black line indicates the best fit line (via the bisector method) to the binned data for all galaxies (see Table~\ref{tab:results}). Filled symbols show bins where the integrated S/N for HCN/CO $\geq 3\sigma$. Downward-pointing arrows show upper limits.
\label{fig:density-density-alma}}
\end{figure*}

\subsection{Average Cloud-Scale CO Intensity in each HCN Beam}
\label{subsec:avgcssd} 

We estimate the cloud-scale surface density from PHANGS-ALMA CO~(2-1) and (for NGC~5194) PAWS CO~(1-0) data at 120~pc resolution (see Figure \ref{fig:fov}). PHANGS ALMA produces CO~(2-1) line maps with $\sim 1''{-}1.5''$ resolution, $2.5$~km~s$^{-1}$ velocity resolution, $\sim 0.1$~K noise per channel, and including short spacing and total power information from ALMA's Morita Atacama Compact Array (ACA). Data reduction and imaging for PHANGS-ALMA are described in A.~K.~Leroy et~al. in preparation. Details regarding the creation of moment maps, noise, and completeness for the four targets studied here appear in \citet{SUN18}. Given the distances and angular resolutions of these data, $120$~pc represents the common physical resolution for our high resolution CO data. We convolve all four CO~(2-1) maps to share this common physical resolution.

For NGC~5194, we also convolve the PAWS CO~(1-0) moment-zero map \citep{SCHINNERER13,PETY13} to 120~pc resolution. To place these measurements on the same CO~(2-1) intensity scale as our other targets, we then multiply the PAWS \mbox{CO~(1-0)} intensities by a typical CO~(2-1)/\mbox{CO~(1-0)} ratio of $0.7$ \citep[uncertain by $\pm0.15$~ dex, see][]{LEROY13}.

We measure HCN/CO at coarser resolution than we measure the CO~(2-1) intensity, $I_{\rm CO(2-1)}$. To connect the two measurements, we calculate the intensity-weighted average $I_{\rm CO(2-1)}$ within each larger HCN beam. This weighted average, $\left< I_{\rm CO(2-1)} \right>$, measures the mean 120~pc resolution $I_{\rm CO(2-1)}$ from which CO photons emerge. Formally,

\begin{eqnarray}
\label{eq:co_21}
\left< I_{\rm CO(2-1)} \right> = \frac{\left( I_{\rm CO(2-1)} \right)^2 * \Omega}{I_{\rm CO(2-1)} * \Omega}~.
\end{eqnarray}

\noindent Here, $I_{\rm CO(2-1)}$ is the CO map at 120~pc resolution, the asterisk represents convolution, and $\Omega$ indicates the Gaussian kernel used to change the resolution of the map from 120~pc to the final resolution (650~pc for the ALMA sample and 2770~pc for the EMPIRE sample). 

$\left< I_{\rm CO(2-1)} \right>$ is the expectation value of CO intensity weighted by itself within each coarser HCN beam. In practice, given some conversion between light and mass (i.e., $\alpha_{\rm CO}$), $\left< I_{\rm CO(2-1)} \right>$ captures the mass-weighted 120~pc resolution surface density of molecular gas inside each larger HCN beam, $\left< \Sigma_{\rm mol} \right>$. 

The advantage of this approach, which is discussed at length by \citet{LEROY16} \citep[and see][]{OSSENKOPF02,UTOMO18} is that it preserves the high resolution information and down-weights empty regions. Compared to the unweighted average, this intensity-weighted average, $\left<  I_{\rm CO(2-1)} \right>$ is $0.6$~dex higher at 650~pc resolution and $1.4$~dex higher at 2770~pc resolution. The intensity weighting is not equivalent to smoothing, as it effectively leverages the high resolution information and yields characteristic surface densities that are $\sim 4{-}30$ times higher than smoothed maps. The difference reflects beam dilution due to the large amount of empty space in the CO maps, which is also visible from Figure \ref{fig:fov}.

\textbf{Uncertainty:} We estimate the uncertainty in $\left< I_{\rm CO(2-1)} \right>$ via a Monte-Carlo calculation. We begin with the original CO~(2-1) cubes, add randomly generated Gaussian noise with the correct mean amplitude, and then run these noise-added cubes through our full analysis procedure. We repeat this process 100 times, and calculate the standard deviation in $\left< I_{\rm CO(2-1)} \right>$ over all realizations. The mean error calculated in this way is ${\sim}2$ K~km~s$^{-1}$. As a result, all $\left< I_{\rm CO(2-1)} \right>$ values in this paper have a signal-to-noise ratio $>5$.

The distances to our targets are uncertain by $\sim 10{-}30\%$. The angular scale corresponding to $120$~pc is correspondingly uncertain, adding an additional uncertainty to our calculation. Using the same data that we use here, \cite{SUN18} showed that changing from a physical resolution of 80~pc to 120~pc (i.e., by $>30\%$) alters the mean $I_{\rm CO(2-1)}$ in a galaxy by $\sim 0.05{-}0.1$~dex.

\subsection{Binned Relation}
\label{subsec:BD}    

Our EMPIRE and ALMA surveys only detect HCN along individual lines of sight at high signal-to-noise in the brightest regions of our targets. However, our data also contain a large amount of information at lower signal-to-noise. 

We recover $\left< I_{\rm CO (2-1)} \right>$ at high signal-to-noise across a wide area. Therefore, to access the fainter HCN emission, we measure the average HCN/CO in bins of $\left< I_{CO (2-1)} \right>$.  We report the integrated HCN divided by the integrated CO in each bin, with the statistical uncertainty on this binned ratio propagated from the original maps following \citet[][]{GALLAGHER18}. This binning increases the signal-to-noise in HCN/CO via averaging and extends the dynamic range in our measured correlation dramatically.

Following \citet{GALLAGHER18}, our HCN integrated intensity maps are created by integrating the cube over the region with bright CO emission, whether or not that region shows HCN emission at high signal-to-noise. As a result, this averaging approach is almost equivalent to spectral stacking using the CO velocity field as a prior \citep[as in][]{SCHRUBA11,JIMENEZDONAIRE17}. We verify this by comparing the two approaches directly. After shifting the HCN cubes to the local mean CO velocity and averaging, we derive stacked line ratios in bins of $\left< I_{\rm CO} \right>$. On average, the spectral stacking yields the same results as our mask-and-average approach within ${\sim}10$\%, with no systematic offset.

\subsection{Interpretation in Terms of Physical Quantities}
\label{subsec:IOQ}

\begin{figure*}[th!]
\plotone{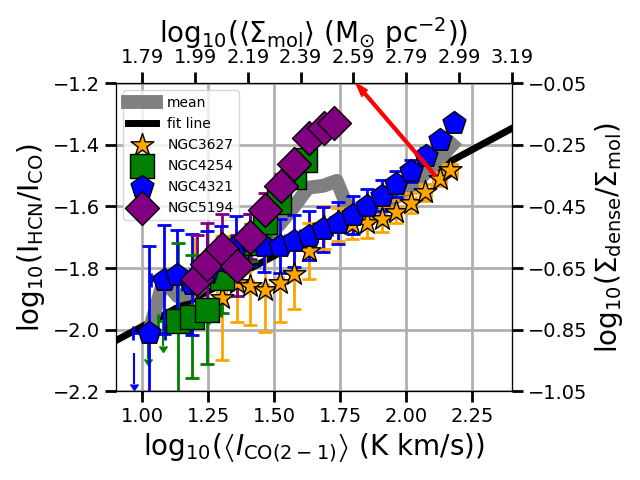}
\caption{As for Figure \ref{fig:density-density-alma}, but now showing results averaged over $\sim 2.8$~kpc regions in the EMPIRE IRAM \mbox{30-m} targets. The red arrow shows the effect of lowering $\alpha_{\rm CO}$ by a factor of $2$ while leaving $\alpha_{\rm HCN}$ constant. Based on \citet{SANDSTROM13} such an adjustment should be appropriate for the inner part of NGC~3627 and NGC~4321 but not NGC~4254 or NGC~5194. \label{fig:density-density-empire}}
\end{figure*}

We report the observed HCN/CO ratio as a function of $\left<I_{\rm CO (2-1)}\right>$. These quantities are interesting because they trace the fraction of dense gas and the mean surface density of molecular clouds. Adopting simple translations from observables, we indicate these two physical quantities on the alternative right and top axes of Figures \ref{fig:density-density-alma} and \ref{fig:density-density-empire}.

To translate HCN/CO to $f_{dense}$, we assume $\alpha_{\rm CO (1-0)} = 4.35$ \acounits\ \citep{BOLATTO13} and a more uncertain $\alpha_{\rm HCN} \approx 14$ \acounits\ to convert HCN to the mass of gas above a density of $n_{\rm H2} \approx 5\times10^3$~cm$^{-3}$ \citep{ONUS18}. For comparison, many previous studies have assumed $\alpha_{\rm HCN} = 10$ \acounits\ for gas above $3 \times 10^4$~cm$^{-3}$ \citep[following][]{GAO04}.

Both the density of gas traced by \mbox{HCN~(1-0)} and the conversion from HCN emission to a dense gas mass remain uncertain. The effective critical density of HCN changes as a function of temperature and optical depth, which are hard to measure \citep{JIMENEZDONAIRE17}. Moreover, gas at densities below the effective critical density still emits HCN, rendering the density traced by HCN a product of the emissivity and density distribution \citep{LEROY17}. For more discussion, see \citet{GAO04}, \citet{USERO15}, \citet{LEROY17}, \cite{ONUS18}, and \citet{GALLAGHER18}, as well as the Milky Way studies by \citet{KAUFFMANN17}, \cite{MILLS17}, and \citet{PETY17}.

Similarly, we report $\left<I_{\rm CO (2-1)}\right>$ as our primary measurement and the 120~pc resolution molecular gas surface density, $\left<\Sigma_{\rm mol}\right>$, as an alternative axis. For a typical CO~(2-1)/\mbox{CO~(1-0)} ratio of $0.7$, the Galactic CO-to-H$_2$ conversion factor is $\alpha_{\rm CO (2-1)} \approx 6.2$~\acounits .

\subsection{Additional Checks}

To check our results, we also analyzed the ALMA HCN data at the EMPIRE common physical resolution of 2770~pc. At a fixed $\left< I_{\rm CO(2-1)} \right>$, the ALMA and EMPIRE data differ by a mean of 15\% in HCN/CO. Mostly, this offset reflects that the two data sets cover different area (see Figure \ref{fig:fov}). When we match the areal coverage (i.e., consider EMPIRE only over the ALMA area), a smaller ${\sim}5\%$ difference remains.


Our CO~(1-0) data (used in the denominator of HCN/CO) come from different sources for EMPIRE and the ALMA sample. We estimate the uncertainty associated with our choice of CO map by considering NGC~4321, for which we have ALMA, BIMA, and IRAM 30-m CO~(1-0) maps. We only have CARMA data for NGC4254 so we cannot explore how CARMA compares to our other CO~(1-0) sources. We repeat our complete analysis with each NGC~4321 CO~(1-0) map. At 650~pc resolution, the HCN/CO ratios measured using the ALMA CO map are $\approx 0.1$~dex lower than those measured using the BIMA map. At 2770~pc resolution, the ALMA map yields HCN/CO ratios $\approx 0.2$~dex lower than BIMA, while the EMPIRE maps yields ratios $\sim 0.25$~dex lower than BIMA. The ALMA and EMPIRE results agree within $0.06$~dex. Overall, we have good confidence in the EMPIRE and ALMA measurements, but the two BIMA-based CO maps may have results uncertain by $\sim 0.1$~dex at 650~pc resolution. While there are offsets between the ratio values calculated using different input data, these offsets do not change the nature of the observed trends.


\section{Results}
\label{sec:results}

\capstartfalse
\begin{deluxetable*}{lccc}[ht!]
\tabletypesize{\scriptsize}
\tablecaption{Results Comparing HCN/CO to $\left< I_{\rm CO(2-1)} \right>$ \label{tab:results}}
\tablewidth{0pt}
\tablehead{
\colhead{HCN Data} & 
\colhead{Resolution} & 
\colhead{Rank.~Corr.} & 
\colhead{Fit Slope, Intercept} 
} 
\startdata 
ALMA & 650~pc & $0.83$ & $0.81 (\pm 0.09), -1.93 (\pm 0.04)$ \\ 
ALMA & 2770~pc & $0.77$ & $0.41 (\pm0.04), -1.73 (\pm 0.02$) \\ 
EMPIRE & 2770~pc & $0.78$ & $0.55 (\pm0.05), -1.73 (\pm 0.01)$ 
\enddata
\tablecomments{Relation between mean 120~pc CO~(2-1) intensity, $\left< I_{\rm CO(2-1)} \right>$, tracing cloud-scale mean surface density, and HCN/CO, tracing $f_{dense}$. {\bf HCN Data}: the source of the HCN data. {\bf Rank Corr.}: Spearman's rank correlation coefficient relating HCN/CO to $\left< I_{\rm CO(2-1)} \right>$ across all bins for all galaxies. All have $p$ values $< 0.01$. {\bf Fit}: Linear fit to the logarithmic data (i.e., power law fit), normalized at the lower end of the $\log_{10} \left< I_{\rm CO(2-1)} \right>$ range for each sample. Shown in the table are the slope ($m$) and intercept ($b$) with associated uncertainties for the following equation: $\log_{10} \frac{HCN}{CO} = m \times \log_{10} \left<I_{\rm CO (2-1)}/(30\rm{K \ km/s})\right> b$. To convert into an approximate relation between $f_{dense}$ and surface density, use $f_{\rm dense} \approx 2.3$~HCN/CO and $\Sigma_{\rm mol} \left[ {\rm M_{\odot}~pc}^{-2} \right] \approx 6.2~\left< I_{\rm CO (2-1)} \right>$.}
\end{deluxetable*}
\capstarttrue

\begin{figure*}
\begin{center}
\includegraphics[width=0.40\textwidth]{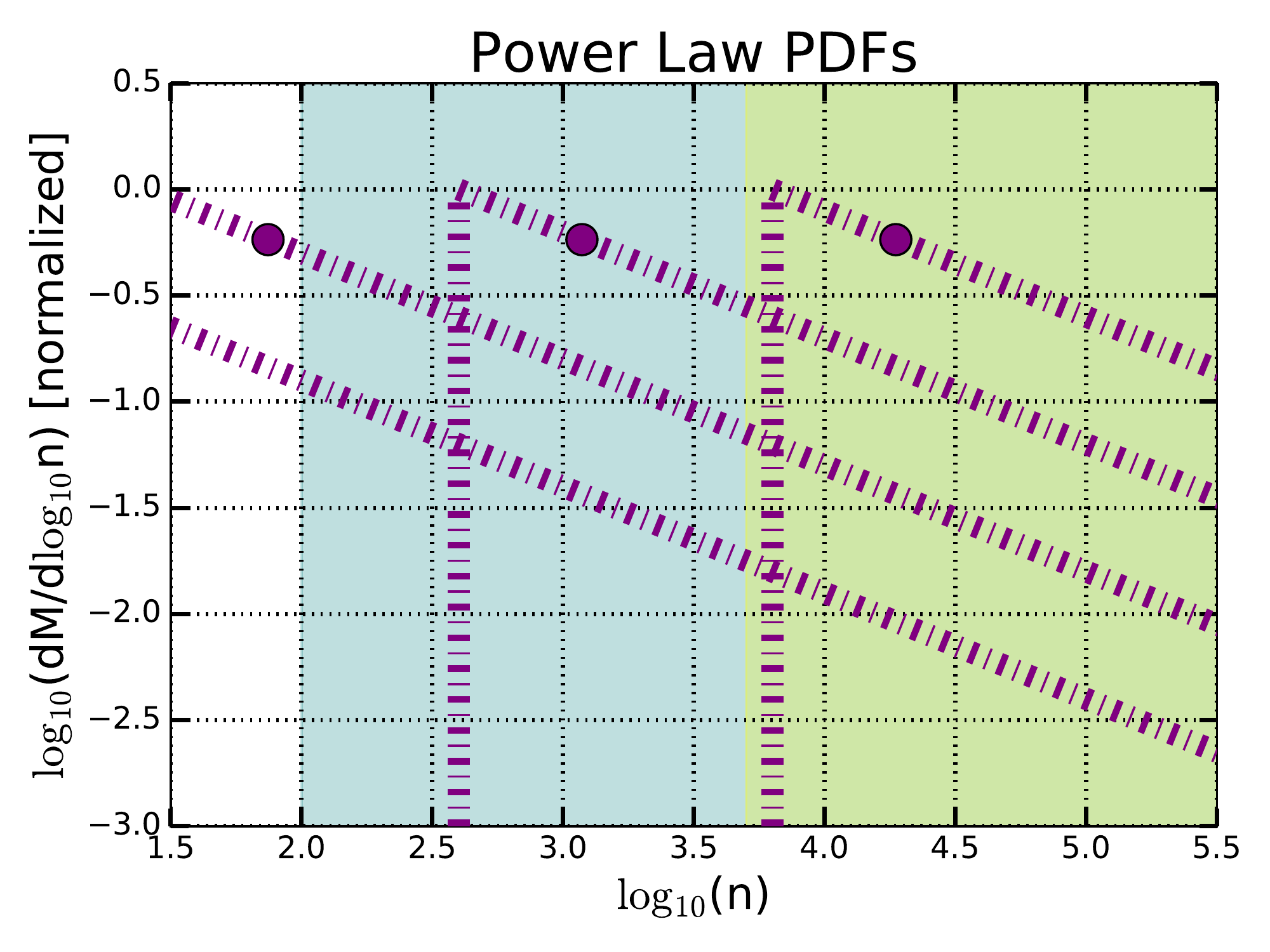}
\includegraphics[width=0.40\textwidth]{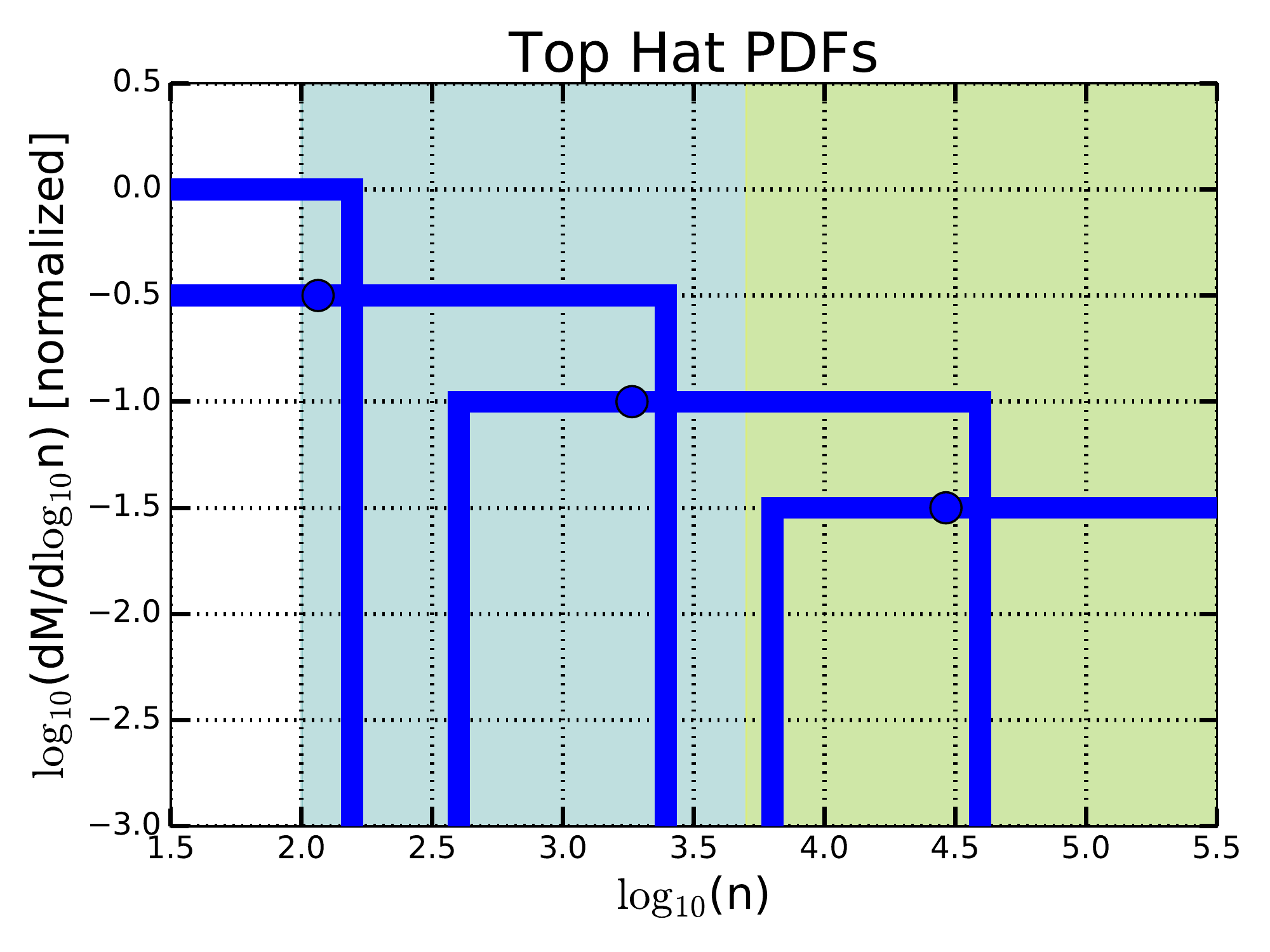}
\includegraphics[width=0.40\textwidth]{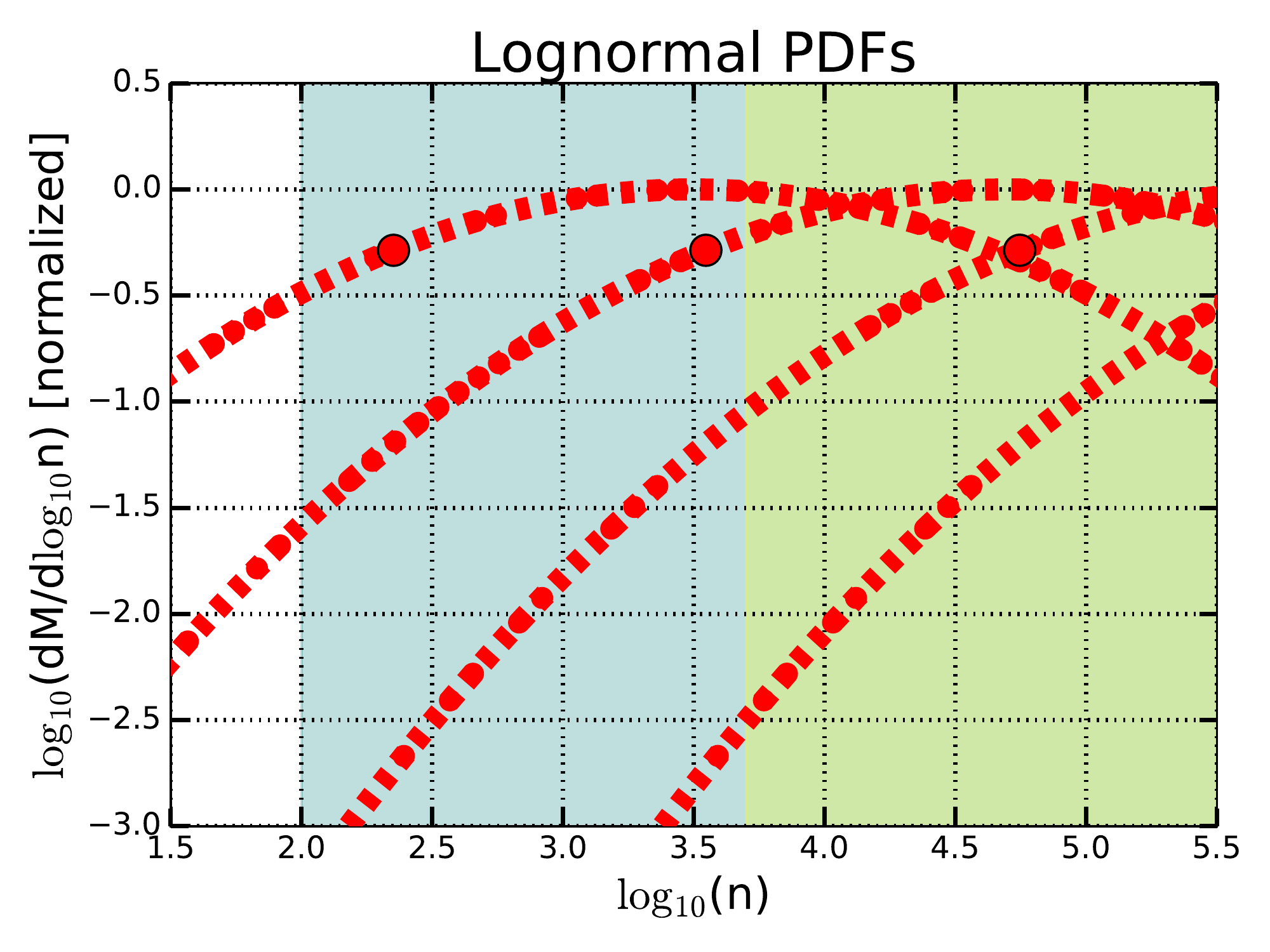}
\includegraphics[width=0.40\textwidth]{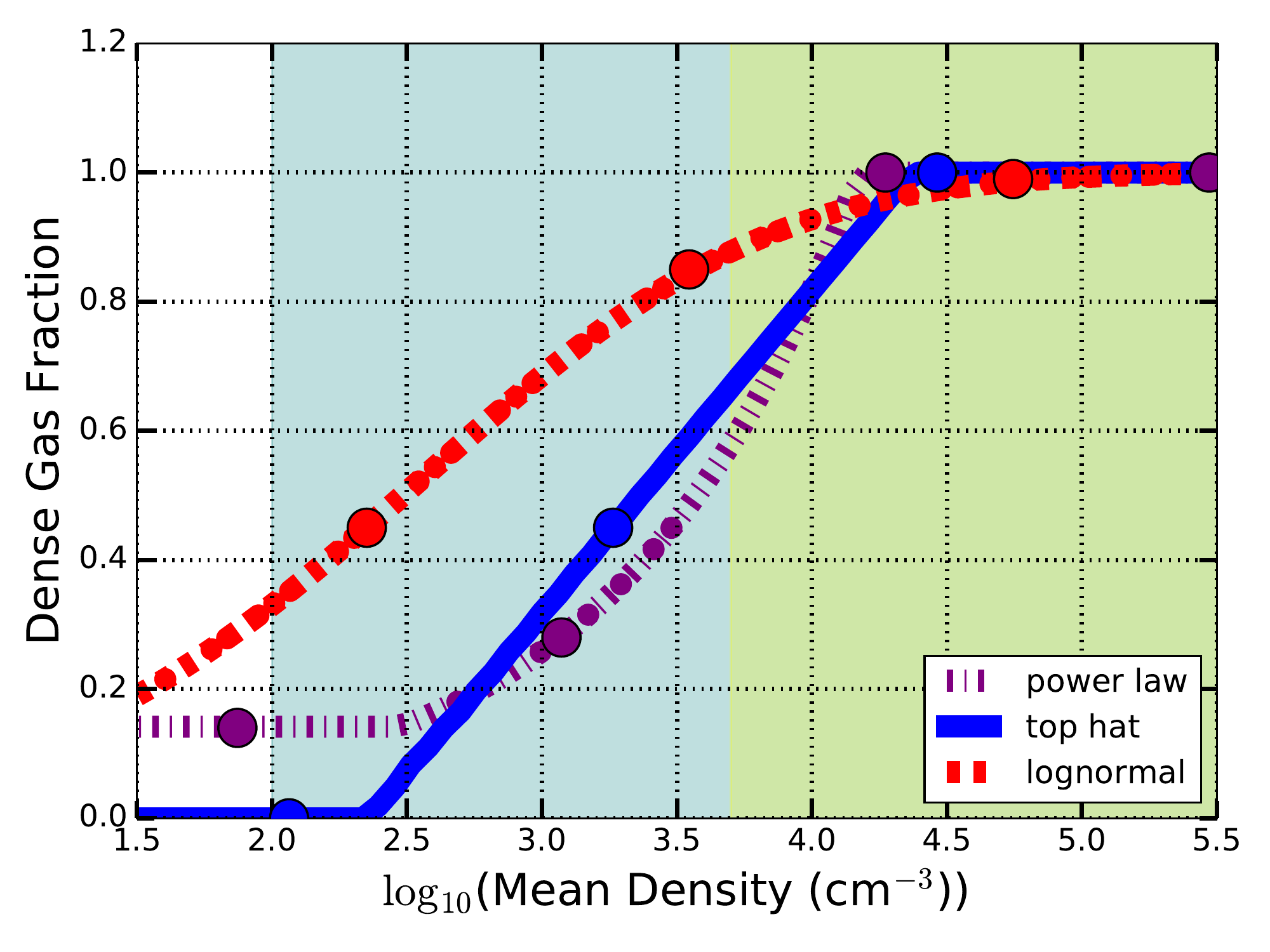}
\end{center}
\caption{Dependence of dense gas mass fraction on mean volume density (bottom right panel) for several distributions of mass as a function of volume density (first three panels). The first three panels illustrate the distribution of mass (with arbitrary normalization) as a function of volume density for our three models: ({\em top left}) a ``bottom heavy'' power law, ({\em top right}) a flat, or ``top hat'' distribution (here the PDFs are vertically staggered for ease of viewing), and ({\em bottom left}) a lognormal. ({\em bottom right}:) The predicted dependence of dense gas mass fraction on the mean (mass weighted) density for each distribution. To calculate $f_{dense}$, we assume a constant emissivity for CO at densities in the teal ($n > 100$~cm$^{-3}$) and green ($n > 5,000$~cm$^{-3}$) shaded regions, and for HCN at densities in the green shaded region. Dividing the mass above the high density threshold (light green) by the mass above the low density threshold (teal) we construct a model ``dense gas fraction'' similar to what we expect to find using the HCN/CO ratio. We calculate the mean density from the expectation value of $n$ across $dp/dn$. The colored points show the mass-weighted mean for each illustrated density distribution.
\label{fig:density-dist}}
\end{figure*}

\subsection{Observed Correlation Between HCN/CO and Cloud-Scale Surface Density}
\label{subsec:OC}

Figures~\ref{fig:density-density-alma} and \ref{fig:density-density-empire} show binned HCN~(1-0)/CO~(1-0) as a function of $\left< I_{\rm CO(2-1)} \right>$ for our two subsamples. HCN~(1-0)/CO~(1-0) and $\left< I_{\rm CO(2-1)} \right>$ are strongly positively correlated for all galaxies in both samples. Spearman's rank correlation coefficient, $\rho$, is high for both individual galaxies ($\rho = 0.97{-}1.0$) and the entire sample ($\rho = 0.77{-}0.83$; Table~\ref{tab:results}). The corresponding low $p$ values indicate that this correlation is unlikely to be produced by random noise. Our measurements offer strong evidence for a significant underlying relationship between our $f_{dense}$ (traced by HCN/CO) and cloud scale surface density (traced by $\left< I_{\rm CO(2-1)} \right>$). 

Table~\ref{tab:results} reports power law fits (fit via the bisector method) that offer a reasonable description of each sample. We plot these as black lines in Figures~\ref{fig:density-density-alma} and \ref{fig:density-density-empire}. 

These fits offer a good first-order description of the observed trends. However, we do observe substantial galaxy-to-galaxy variations in HCN/CO at fixed $\left< I_{\rm CO(2-1)} \right>$. The standard deviation in HCN/CO at fixed $\left< I_{\rm CO(2-1)} \right>$ is 0.11~dex for the ALMA sample and 0.08~dex for the EMPIRE sample. Because of our binning approach, this reflects \textit{only} the galaxy-to-galaxy scatter. If we were in a position to measure the cloud-to-cloud or region-to-region scatter within each $\left< I_{\rm CO(2-1)} \right>$, we would expect to find more variation.

NGC~3627 and NGC~4321 in the EMPIRE data (Figure \ref{fig:density-density-empire}) also exhibit different behavior at high and low $\left< I_{\rm CO (2-1)} \right>$. For these galaxies, the slope relating HCN/CO to $\left< I_{\rm CO(2-1)} \right>$ steepens near $\log_{10} \left< I_{\rm CO(2-1)} \right> \sim 1.75$ ($\sim 55$~K~km~s$^{-1}$ $\sim 350$~M$_\odot$) and these galaxies show higher $\log_{10} \left< I_{\rm CO(2-1)} \right>$ and lower HCN/CO compared to the other two targets. Though we do not find similar curvature, the same two targets show a similar offset from NGC4254 to high $\log_{10} \left< I_{\rm CO(2-1)} \right>$ and lower HCN/CO in the ALMA observations (Figure \ref{fig:density-density-alma}). Below, we suggest that ``starburst''- like conversion factors in the centers of these galaxies offer a likely explanation for this behavior.

\subsection{Implications}
\label{subsec:TI}

The correlation between HCN/CO and $\left< I_{\rm CO (2-1)} \right>$ supports the idea that both quantities trace the density distribution of molecular gas. This suggests that the mean surface density of a molecular cloud and its dense gas content both reflect an underlying, environment-dependent gas density distribution.

Below, we show that this would be expected from simple models as long as cloud-scale mean \textit{surface} density traces cloud-scale mean \textit{volume} density. In fact, as discussed in \citet[][see their Figure 5]{LEROY17b}, surface and volume density do correlate in recent molecular cloud catalogs. Moreover, the molecular gas scale height in the Milky Way appears relatively constant over the inner $\sim 8$~kpc \citep{HEYER15}. In short, current evidence appears to support the idea that at high resolution molecular gas surface density tracks molecular gas volume density to first order \citep[see also][]{UTOMO18}.

\textbf{Connection to Environment:} The regions of galaxies with high gas and stellar surface densities and high interstellar gas pressure also tend to have high dense gas fractions, $f_{dense}$ \citep{USERO15,BIGIEL16,GALLAGHER18}. These also tend to be in the inner parts of galaxies, so the binned results in Figures \ref{fig:density-density-alma} and \ref{fig:density-density-empire} also roughly map to radius and stellar surface density, with the central regions of each galaxy mostly contributing to the top right part of the relationship.

At the same time, many recent studies have shown an environmental dependence of the cloud-scale properties of molecular gas \citep[e.g.,][]{HUGHES13, COLOMBO14, LEROY16, SUN18, FAESI18}. Broadly, these results have the same sense as those for $f_{dense}$. The internal pressure of molecular clouds appears to correlate with large scale environmental gas pressure, radius, and the stellar mass of the host galaxy \citep[][]{HUGHES13,SUN18,SCHRUBA18}.

Our results directly connect these two lines of evidence. The correlation that we observe suggests that the properties of the bulk molecular gas \citep[e.g.,][]{HUGHES13, COLOMBO14, LEROY16, SUN18, FAESI18, SCHRUBA18} and $f_{dense}$ \citep{GAO04,USERO15,BIGIEL16,GALLAGHER18} reflect different aspects of the same environment-dependent density distribution. The sense of this correlation should broadly be that high pressure, high surface density, inner parts of galaxies have both high $f_{dense}$ and high cloud-scale mean surface density.

\textbf{Does $\alpha_{\rm CO}$ Drive Galaxy-to-Galaxy Variations?} \cite{SANDSTROM13} found that the CO-to-H$_2$ conversion factor ($\alpha_{\rm CO}$) is often lower than the standard Galactic value in the inner regions of galaxies with dense, bar-fed centers. Specifically, they found $2{-}3$ times lower $\alpha_{\rm CO}$ in the center of NGC~3627 and NGC~4321 compared to the disks. NGC~4254 \citep{SANDSTROM13} and NGC~5194 \cite{LEROY17b} do not show central $\alpha_{\rm CO}$ depressions, and NGC~4254 may even show a central rise in $\alpha_{\rm CO}$. An arrow in Figure \ref{fig:density-density-empire} indicates the effect of lowering $\alpha_{\rm CO}$ (but not $\alpha_{\rm HCN}$) by a factor of $2$. Adjusting the inner (high $\left< I_{\rm CO (2-1)} \right>$) NGC~3627 and NGC~4321 points in this way would bring the different galaxies into better agreement. If $\alpha_{\rm CO}$ and $\alpha_{\rm HCN}$ change in the same way, then the points only move horizontally.

Though not a unique explanation, a low $\alpha_{\rm CO}$ due to bright diffuse (but not dense) molecular gas offers a feasible explanation for some of the offset among our targets. This change in $\alpha_{\rm CO}$ as a function of $\left< I_{\rm CO (2-1)} \right>$) could also explain the curvature in the EMPIRE results for NGC~3627 and NGC~4321. 


\textbf{Indirect Evidence That Both Axes Trace Density:} A number of recent studies have raised concerns about the ability of HCN to trace dense gas \citep[e.g.,][]{PETY17, LEROY17, KAUFFMANN17, RATHBORNE15}. The unknown abundance of the molecule, its uncertain opacity \citep{JIMENEZDONAIRE17}, and possible excitation effects \citep{SHIMAJIRI17} all are likely to affect the HCN-to-dense gas conversion factor. Similarly, $\left< I_{\rm CO (2-1)} \right>$ at 120~pc resolution may suffer from excitation effects \citep{KODA12}. Variations in the line of sight depth could also introduce scatter in the relationship between mean volume density and mean surface density. 

Despite these concerns, these two observables remain among our most practical tracers of $f_{dense}$ and mean cloud-scale surface density for observations of external galaxies. Our finding that they track each other is a powerful, though still indirect, evidence that both the HCN/CO ratio and the cloud-scale CO intensity are meaningful tracers of gas density. FurtherMoreover, even if HCN traces lower density gas than is commonly assumed, the contrast between the HCN and CO lines still captures the shape of the density distribution to some extent.

\subsection{Expectations From Simple Models}
\label{subsec:SM}

Our observed correlation would be expected if (1) HCN/CO traces the fraction of gas above some threshold density, (2) $\left< I_{\rm CO(2-1)} \right>$ traces the mean surface density of molecular clouds, and (3) the mean surface density of a molecular cloud traces its mean volume density. Here we illustrate this by integrating over some simple models of gas density distributions.

We calculate $f_{dense}$ as a function of mean density for three model gas volume density distributions\footnote{In our notation $p$ refers to the probability of finding a given piece of volume in the cloud to have density $n$. $dp/dn$ refers to the probability density distribution as a function of density. The mass of material at a density $n$ will be $m = n \times p$. In figures we plot $\log d m / d \log n$, illustrating the distribution of mass as a function of density.}: (1) a ``bottom heavy'' power law with a slope of $-2.5$, such that $dp/dn \propto n^{-2.5}$; (2) a power law with $dp/dn \propto n^{-2}$ and a width of 2~dex; such a distribution has equal mass per logarithmic bin (i.e., a ``top hat'' in mass); and (3) a lognormal distribution with $1\sigma$ width of $1.0$~dex. Power law and lognormal distributions are currently the most popular ways to represent volume and column density distributions \citep[e.g.,][]{PADOAN02, KAINULAINEN09, LOMBARDI15}.

Figure~\ref{fig:density-dist} shows the shape of each model distribution for several mean densities. To calculate the dense gas mass fraction, we take the ratio of mass above two thresholds, one at low density (teal) and one at high density (light green). These thresholds correspond to the $n_{eff}$ of the two lines. Dividing the mass above the high density threshold by the mass above the low density threshold we construct a model ``dense gas fraction'' similar to what we expect to find using the HCN/CO ratio.

To predict how $f_{dense}$ depends on mean density, we repeat the exercise for many distributions. We leave the width (when applicable) and slope of each distribution fixed, keep the CO and HCN density thresholds fixed, and vary only the mean density of the cloud. The lines in the bottom right panel of Figure~\ref{fig:density-dist} shows the resulting $f_{dense}$ as a function of mean density (the points indicate the specific cases illustrated in the other panels). 

For all distributions, Figure~\ref{fig:density-dist} shows that as the mean density increases, a larger and larger fraction of the gas sits above the effective density of \mbox{HCN~(1-0)} (the light green region). As a result, we expect a correlation between mean density and $f_{dense}$. At intermediate densities, where the mean density lies between the low (CO) and high (HCN) density threshold, power-law-like scaling relations between mean density and $f_{dense}$ are common. If our observed $\left< I_{\rm CO (2-1)} \right>$, tracing the mean cloud-scale \textit{surface} density, also traces the mean cloud-scale \textit{volume} density then our observations would match the expectation from these simple models.

\section{Summary}
\label{sec:summary}

We measure how the ratio of \mbox{HCN~(1-0)} to \mbox{CO~(1-0)} emission depends on the 120~pc scale CO intensity, $\left< I_{\rm CO(2-1)} \right>$, in five nearby galaxies. HCN/CO traces the dense gas mass fraction, while $\left< I_{\rm CO(2-1)} \right>$ measures the cloud-scale molecular gas mass surface density.

We find a strong correlation between these two quantities, albeit with differences in the shape and normalization from galaxy to galaxy. This could be expected if these two quantities trace different aspects of the same underlying distribution of gas densities. We illustrate this using simple model density distributions.

This result supports a view in which the large-scale structure of a galaxy shapes the local gas density distribution. Both the mean cloud-scale gas surface density, which is often measured as a property of molecular clouds, and the dense gas fraction, which is probed via spectroscopy, reflect this distribution. In this case, recent results tracing the environmental dependence of molecular cloud properties \citep[e.g.,][]{HUGHES13,COLOMBO14,SUN18} and those showing a dependence of dense gas fraction on local disk structure \citep[e.g.,][]{USERO15,BIGIEL16,GALLAGHER18} capture highly related aspects of the coupling between the physical state of cold gas and galactic environment.

Our analysis only scratches the surface of what can be done comparing cloud properties to density sensitive spectroscopy. In the near future, it should be possible to expand this sample by combining PHANGS-ALMA CO maps with new HCN observations from ALMA, the Green Bank Telescope, and the IRAM \mbox{30-m}. Expanding the analysis to a suite of lines with a wide range of critical densities will better constrain the volume density distribution \citep{LEROY17,GALLAGHER18}. The CO imaging also includes information on cloud-scale dynamics, which will allow us to test how the the cloud-scale velocity dispersion and virial parameter relate to gas density and the star formation. Finally, as mentioned above, our knowledge of how HCN and similar lines trace dense gas is improving rapidly thanks to ongoing theoretical \citep[e.g.,][]{ONUS18}, extragalactic \citep[e.g.,][]{JIMENEZDONAIRE17}, and Galactic studies \citep[e.g.,][]{PETY17,KAUFFMANN17,MILLS17}.

\acknowledgments We thank the anonymous referee for a fast and constructive report that improved the quality of the paper. This paper makes use of the following ALMA data: ADS/JAO.ALMA \#2015.1.00956.S, ADS/JAO.ALMA \#2013.1.00634.S, ADS/JAO.ALMA \#2011.0.00004.SV. ALMA is a partnership of ESO (representing its member states), NSF (USA) and NINS (Japan), together with NRC (Canada) and NSC and ASIAA (Taiwan), and KASI (Republic of Korea), in cooperation with the Republic of Chile. The National Radio Astronomy Observatory is a facility of the National Science Foundation operated under cooperative agreement by Associated Universities, Inc. The Joint ALMA Observatory is operated by ESO, AUI/NRAO and NAOJ. MG acknowledges generous support from the NRAO student observing support program. The work of MG, AKL, JS, and DU is partially supported by the National Science Foundation under Grants No. 1615105, 1615109, and 1653300. FB acknowledges funding from the European Union’s Horizon 2020 research and innovation programme (grant agreement No 726384). ES, CF, and TS acknowledge funding from the European Research Council (ERC) under the European Union’s Horizon 2020 research and innovation programme (grant agreement No. 694343). AU acknowledges support from Spanish MINECO grants ESP2015-68964 and AYA2016-79006. SCOG acknowledges support from the DFG via SFB 881 ''The Milky Way System'' (sub-projects B1, B2 and B8). MRK acknowledges support from the Australian Research Council (Discovery Projects award DP160100695, and the Centre of Excellence for All Sky Astrophysics in 3 Dimensions, project CE170100013). JMDK and MC gratefully acknowledge funding from the German Research Foundation (DFG) in the form of an Emmy Noether Research Group (grant number KR4801/1-1). JMDK gratefully acknowledges funding from the European Research Council (ERC) under the European Union’s Horizon 2020 research and innovation programme via the ERC Starting Grant MUSTANG (grant agreement number 714907). DC is supported by the European Union's Horizon 2020 research and innovation programme under the Marie Sk\l{}odowska-Curie grant agreement No 702622. JP acknowledges support from the Program National "Physique et Chimie du Milieu Interstellaire" (PCMI) of CNRS/INSU with INC/INP, co-funded by CEA and CNES. ER acknowledges the support of the Natural Sciences and Engineering Research Council of Canada (NSERC), funding reference number RGPIN-2017-03987.

\end{document}